\documentclass[conference]{IEEEtran}
\usepackage{cite}
\usepackage{amsmath,amssymb,amsfonts}
\usepackage{algorithmic}
\usepackage{graphicx}
\usepackage{textcomp}
\usepackage{xcolor}
\def\BibTeX{{\rm B\kern-.05em{\sc i\kern-.025em b}\kern-.08em
    T\kern-.1667em\lower.7ex\hbox{E}\kern-.125emX}}
\usepackage{bm}
\usepackage{color}
\usepackage{balance}
    
\newcommand{\bc}[1]{\mbox{\boldmath $\mathcal{#1}$}}
\newcommand{\mb}[1]{\mathbf{#1}}

\newcommand{\T}{\mathrm{T}}

\newenvironment{mat}[1]{\left[\begin{array}{#1}}{\end{array}\right]}
\newcommand{\bmx}[1]{\begin{mat}{#1}}
\newcommand{\emx}{\end{mat}}
    
\begin{document}

\title{Beyond Tensor Probabilistic Independent Component Analysis --- Putting Block-Term Decomposition and Independent Vector Analysis Together}

\author{\IEEEauthorblockN{Eleftherios Kofidis}
\IEEEauthorblockA{\textit{Dept. of Statistics and Insurance Science} \\
\textit{University of Piraeus}\\
185~34 Piraeus, Greece \\
Email: kofidis@unipi.gr}
}

\maketitle

\begin{abstract}
Tensor probabilistic independent component analysis (TPICA) is a popular approach to analyzing functional magnetic resonance imaging (fMRI) data, which draws its popularity from its ability to enrich the advantages of the statistics-based ICA with the awareness of the multi-way nature of these data, brought about and exploited via a deterministic 3-way (time $\times$ space $\times$ subjects) tensor decomposition (Canonical Polyadic Decomposition (CPD)) model. It has, however, received critique concerning its robustness in realistic fMRI unmixing scenarios, notably those involving sources that are strongly overlapped in space. Such cases may not meet the assumption of statistical independence required in ICA. They can instead be better described as independent vectors (or subspaces) of dependent components, pointing to the adoption of alternative statistical approaches, notably independent vector analysis (IVA). On the other hand, on the deterministic side, CPD is often restrictive and is outperformed by the more flexible block-term decomposition (BTD) model, also in the fMRI source unmixing context. Given the above, plus strong evidence of links between IVA and BTD, it is deemed worthwhile to consider the possibilities of generalizing TPICA to a BTD-based ``TPIVA" extension, which would more successfully combine the power of statistics and tensor decomposition. This could also entail a generalization of the BTD model, where (non)collinearity would be replaced by statistical (in)dependence. This note aims to outline the state-of-the-art and the above ideas in more detail, serving as a preliminary, motivating step in this research direction.
\end{abstract}

\begin{IEEEkeywords}
Block-Term Decomposition (BTD), Canonical Polyadic Decomposition (CPD), Independent Vector Analysis (IVA), Probabilistic independent component analysis (PICA), Tensor PICA (TPICA)
\end{IEEEkeywords}

\section{Introduction}
\label{sec:intro}

Besides the successful application of independent component analysis (ICA)-based methods in data-driven functional magnetic resonance imaging (fMRI) data analysis, the inherently multi-way nature of brain imaging data has also inspired the development of the tensor-based approach~\cite{sdfhpf17} to address fMRI analysis problems~\cite{l08}, such as blind source unmixing~\cite{ar04} and functional connectivity quantification~\cite{mcm17}. Following a study of the application of probabilistic ICA (PICA) in fMRI analysis~\cite{bs04}, which featured the estimation of the data dimensionality (number of sources of activation) via probabilistic principal component analysis (PPCA) and the use of prior information, Beckman and Smith~\cite{bs05} went ahead to enrich PICA with the awareness of the multi-way nature of fMRI data, brought about and formalized via a deterministic 3-way (time $\times$ space $\times$ subjects) tensor decomposition (canonical polyadic decomposition (CPD)) model. The resulting tensor probabilistic independent component analysis (TPICA) method has shown advantages over both its PICA and CPD constituents in multisubject/multisession fMRI analysis tasks~\cite{bs05}. 


This otherwise popular approach\footnote{A Python implementation of TPICA is found in the Multivariate Exploratory Linear Optimized Decomposition into Independent Components (MELODIC) tool (https://fsl.fmrib.ox.ac.uk/fsl/fslwiki/MELODIC) of the FMRIB Software Library) (https://fsl.fmrib.ox.ac.uk/fsl/fslwiki/FSL), developed by the Wellcome Centre for Integrative Neuroimaging (WIN) of the University of Oxford, UK, and adopted in more than~1000 labs worldwide.} has, however, received critique concerning its robustness in realistic fMRI unmixing scenarios, notably those involving sources that are strongly overlapped in space. Such cases may not meet the assumption of statistical independence underlying PICA sufficiently well. Thus, \cite{s07}, followed by \cite{hh13}, have instead proposed the adoption of deterministic CPD, which is by definition robust to statistical model mismatches. In fact, \cite{hh13} has even questioned a series of published studies that were based on the assumed advantages of TPICA. Nevertheless, CPD fails to capture states characterized by rich spatial content, due to its strict requirement of rank-1 structure for the data components. This was made explicit in~\cite{ckmt19}, where the block-term decomposition (BTD) model was instead demonstrated to be the most appropriate one for such realistic scenarios. It was a 4-way BTD model, resulting from a tensorization followed by partial unfolding of the spatial mode, that proved to be the best compromise between source unmixing/localization performance and complexity~\cite{ckmt19}. 

In addition to the issue of its sensitivity to source overlapping, ICA has been also criticized as being susceptible to noise (a non-negligible issue, given the commonly low signal-to-noise ratio (SNR)~\cite{l08} in fMRI data\footnote{BTD has been demonstrated to enjoy increased robustness to noise (cf., e.g., \cite{ckmt19}).}) and artifacts~\cite{l08}. Moreover, ICA methods mostly rely on the assumption of an invariant hemodynamic response function (HRF), which is known to be quite unrealistic both subject- and voxel-wise. Such inaccuracies can be better coped with via independent vector analysis (IVA)~\cite{kll06}, which can also resolve the intrinsic permutation ambiguity problem. \cite{lljy08} has demonstrated performance gains for IVA over TPICA, explained by the fact that, in contrast to TPICA, which assumes the same time course (with subject-specific weights) across subjects, the IVA model allows individual-specific time courses that better match the reality of variable HRF. 

Given the above, plus the existence of links between IVA (and its independent subspace analysis (ISA) generalization~\cite{t06}) and BTD~\cite{lahat21}, it is deemed worthwhile to consider possibilities of generalizing TPICA to a BTD-based ``TPIVA" method that would more successfully combine the power of the statistics and tensor decomposition worlds, alleviating their limitations and merging their merits. This note aims to discuss some ideas in this research direction. 

\section{Tensor Probabilistic Independent Component Analysis (TPICA)}
\label{sec:TPICA}

Consider the fMRI analysis problem as stated in~\cite{bs05}, namely having $K$ datasets (corresponding to $K$ subjects or sessions), each with $I$ voxels measured throughout $J$ time instants. Then, under the linear mixing model (LMM), the $I\times J\times K$ fMRI signal tensor\footnote{Note that the number of voxels, $I$, is expected to be much larger than the number of time instants, $J$, and the number of subjects/sessions, $K$. This tensor formulation, therefore, results in an unbalanced 3-way array. This problem could be mitigated by further folding the spatial mode, as in~\cite{ckmt19}.} will obey the following CPD
\begin{equation}
    \bc{X}=\sum_{r=1}^{R}\mb{a}_r\circ\mb{b}_r\circ\mb{c}_r+\bc{E}=[\![\mb{A},\mb{B},\mb{C}]\!]+\bc{E},
    \label{eq:CPD}
\end{equation}
where $\bc{E}$ is the noise/modeling error tensor (assumed white Gaussian\footnote{In the fMRI space, the noise is Rice-distributed, and one can examine alternative ways of model fitting that take this into account (see, e.g., \cite{cvktlv19}). Nonetheless, in high SNR regions, noise is approximately Gaussian~\cite{l08}.}) and the $[\![\cdot]\!]$ Kruskal operator notation for CPD has been used.
 With $R$ sources, the $I\times R$ matrix $\mb{A}=\left[\begin{array}{cccc} \mb{a}_1 & \mb{a}_2 & \cdots & \mb{a}_R \end{array}\right]$ has the corresponding spatial maps at its columns, the time-courses are at the $R$ columns of $\mb{B}=\left[\begin{array}{cccc} \mb{b}_1 & \mb{b}_2 & \cdots & \mb{b}_R \end{array}\right]$, and the $K\times R$ matrix $\mb{C}=\left[\begin{array}{cccc} \mb{c}_1 & \mb{c}_2 & \cdots & \mb{c}_R \end{array}\right]$ contains the subject-specific weights. 

Take the $r$th component in the above CPD model, $\mb{a}_r\circ\mb{b}_r\circ\mb{c}_r$. The first outer product signifies that all $I$ voxels have the same time course, $\mb{b}_r$, modulated with their activation weights, $\mb{a}_r$. What the second outer product implies is that this spatiotemporal mixing is repeated in all subjects (or sessions), with possibly different scaling, found in $\mb{c}_r$. Neglecting the noise for the simplicity of the presentation, and matricizing the tensor with respect to its 1st mode and transposing the resulting $I\times JK$ matrix yields
\begin{equation}
\mb{X}_{(2)}^{\T} = (\mb{C}\diamond \mb{B})\mb{A}^{\T}\triangleq \mb{M}\mb{A}^{\T},
\label{eq:X=MB}
\end{equation}
with $\diamond$ denoting the Khatri-Rao (KR) product. This is the well-known LMM, which turns into the ICA model if the assumption of statistically independent spatial maps (rows of $\mb{A}^{\T}$) is also made. TPICA thus consists of the following two steps. First, apply some ICA algorithm to estimate the sources $\mb{A}$ and the mixing $\mb{M}$. Second, project the latter onto the set of Khatri-Rao product matrices. This can be done by approximating each column of the $\mb{M}$ estimate by a Kronecker product (rank-1 approximation of the matricized column), with the aid of singular value decomposition (SVD), a procedure known as Khatri-Rao least-squares factorization. \cite{bs05} suggests iterating these two steps until convergence (back-projecting to refine the source estimates, PICA-ing them to refine the mixing matrix estimate, decomposing the latter as a KR product, etc.). This procedure amounts to an alternating least squares (ALS) one with projections of the $\mb{A}$ factor onto the set of matrices with statistically independent columns (effected with the aid of PICA). The practice of TPICA shows that no more than two iterations are needed, while the authors of~\cite{hh13} argue that iterating these steps makes no sense 
and instead propose to execute the TPICA steps only once. It should be emphasized that, however the separation is done, via TPICA with or without iterations or via direct CPD, one must also estimate the tensor rank, $R$, that is, the number of sources.\footnote{An NP-hard problem in general~\cite{sdfhpf17}.} The critics of TPICA~\cite{s07,hh13} demonstrate deterministic CPD to be superior in scenarios with spatially overlapped sources only when $R$ has been correctly determined or is \emph{a priori} known. CPD is known to be sensitive to rank over-estimation, however. The model order selection problem, of fundamental importance in obtaining reproducible and explainable results~\cite{akasca22}, appears not to have been given a satisfactory solution in the tensorial fMRI literature yet (cf.~\cite{w12} for an overview of Bayesian and other approaches to this problem).

\noindent
\emph{Remarks.}
\begin{enumerate}
    \item The authors of~\cite{bs05} also view their approach as a \emph{multilevel} version of the classical (known regressor-based) general linear model (GLM) fMRI analysis framework. In the latter, the subject-specific GLM parameters obey a GLM model as well, which involves group-wise parameters. Combining the GLM levels may be seen to lead to a multilinear relation. This should be kept in mind together with the existence of non-blind tensor-based tools for neuroimaging data analysis~\cite{zlz13}.
    \item The not unrealistic scenario of different subjects having similar contributions to spatially different signals is given special mention in~\cite{bs05} and corresponds to a model as in~\eqref{eq:CPD} with (close to) collinear columns in $\mb{C}$. There may also exist groups of (close to) collinear time courses (columns of $\mb{B}$). In such cases, the tensor decomposition model becomes a \emph{degenerate} CPD, which has been studied in~\cite{s12}, among others. Modifications to ALS for coping with such scenarios, such as the coherence-constrained ALS method of~\cite{fgc18}, are worth mentioning here. The way \cite{s12} treats such cases is by fitting a Tucker decomposition (TKD) model instead, constrained to have a block-diagonal core tensor. This is nothing but the so-called block-term decomposition (BTD) model, detailed next. 
\end{enumerate}

\section{Block-Term Decomposition (BTD)}
\label{eq:BTD}

In its general form, BTD~\cite{ldl08a,ldl08b,ldl08c} is stated as decomposing the tensor into a sum of tensors that are Tucker decomposed. Equivalently, this is a single Tucker decomposition, with a block diagonal core. It can thus be seen as an intermediate between the Tucker and the canonical polyadic decomposition. To re-use the analogy made in~\cite{ldl12}, CPD decomposes a tensor into atoms, whereas BTD does so into molecules. To see this, take the particular yet most commonly adopted BTD model of decomposition in rank-$(L_r,L_r,1)$ terms:
\begin{equation}
    \bc{X}=\sum_{r=1}^{R}(\mb{A}^{(1)}_r\mb{A}^{(2)\T}_r)\circ \mb{b}_r,
    \label{eq:LL1}
\end{equation}
commonly referred to as the LL1 model, where the factors $\mb{A}^{(1)}_r,\mb{A}^{(2)}_r$ are of (full column) rank $L_r$, $r=1,2,\ldots,R$. Rank-$R$ CPD results as a special case with all $L_r$ set to unity. BTD has been proven to be the natural tensor model for linear mixing processes involving low-rank mixing and/or source components, in general~\cite{bdl17} and fMRI~\cite{ckmt19} in particular. In the latter application, and by folding the spatial mode into two modes (e.g., $(x,y,z)$ to $(xy,z)$), rank-$(L_r,L_r,1)$ BTD shows up to be well-suited to represent a single-subject space $\times$ time fMRI signal, capturing sources with spatial maps of rank $L_r>1$, which is the case for practically all sources of activation appearing in practice. Such scenarios can also be represented by CPD, but its rank, $R$, would have to be significantly larger, and the extracted components would not be easy to interpret. One can see this by expanding the sum in~\eqref{eq:LL1},
\begin{eqnarray}
    \bc{X} & = & \sum_{l=1}^{L_1}\mb{a}^{(1)}_{1,l}\circ\mb{a}^{(2)}_{1,l}\circ\mb{b}_1+\sum_{l=1}^{L_2}\mb{a}^{(1)}_{2,l}\circ\mb{a}^{(2)}_{2,l}\circ\mb{b}_2+ \nonumber \\ 
    & & \cdots+\sum_{l=1}^{L_R}\mb{a}^{(1)}_{R,l}\circ\mb{a}^{(2)}_{R,l}\circ\mb{b}_R,
    \label{eq:LL1expanded}
\end{eqnarray}
whereby a polyadic decomposition of $\sum_{r=1}^{R}L_r$ rank-1 terms results. This can be equivalently expressed as a Tucker decomposition, namely\footnote{$\times_n$ stands for mode-$n$ product.}
\[    
\bc{X}=\bc{G}\times_1\mb{A}^{(1)}\times_2\mb{A}^{(2)}\times_3\mb{B},
\]
where the $\sum_{r=1}^{R}L_r\times\sum_{r=1}^{R}L_r\times R$ core tensor $\bc{G}$ has its $r$th frontal slice equal to $\mathrm{blockdiag}(\mb{0}_{\sum_{q=1}^{r-1}L_r},\mb{1}_{L_r},\mb{0}_{\sum_{q=r+1}^{R}L_q})$ and $\mb{A}^{(i)}=\bmx{cccc} \mb{A}^{(i)}_1 & \mb{A}^{(i)}_2 & \cdots & \mb{A}^{(i)}_R \emx$, $i=1,2$, and $\mb{B}=\bmx{cccc} \mb{b}_1 & \mb{b}_2 & \cdots & \mb{b}_R \emx$. It is readily verified that $\bc{G}=\bc{I}_{3,\sum_{r=1}^{R}L_r}\times_3\mathrm{blockdiag}(\mb{1}_{L_1},\mb{1}_{L_2},\ldots,\mb{1}_{L_R})^{\T}$, with $\bc{I}_{3,\sum_{r=1}^{R}L_r}$ being the identity tensor, and hence the above can also take the form of the following CPD
\begin{equation}
    \bc{X}=\bc{I}_{3,\sum_{r=1}^{R}L_r}\times_1\mb{A}^{(1)}\times_2\mb{A}^{(2)}\times\mb{\bar{B}},
    \label{eq:LL1CPD}
\end{equation}
with 
\[
\mb{\bar{B}}=\bmx{cccc} \underbrace{\mb{b}_1\;\; \cdots \;\; \mb{b}_1}_{L_1\ \mbox{times}} & \underbrace{\mb{b}_2\;\; \cdots \;\; \mb{b}_2}_{L_2\ \mbox{times}} & \cdots & \underbrace{\mb{b}_R\;\; \cdots \;\; \mb{b}_R}_{L_R\ \mbox{times}}\emx
\]

The CPD equivalent of BTD described above may explain the large values for the number of sources estimated in several works, including~\cite{bs05}. There are many fewer, with similar subject weights (or in other modes, such as similar time courses~\cite{bs05}), that a CPD modeling wrongly views as distinct. 
In the multi-subject case, the corresponding 4-way (space $\times$ space $\times$ time $\times$ subject) tensor is represented by a rank-$(L_r,L_r,1,1)$ BTD model and~\eqref{eq:LL1expanded} takes the form
\begin{eqnarray}
    \bc{X} & = & \sum_{l=1}^{L_1}\mb{a}^{(1)}_{1,l}\circ\mb{a}^{(2)}_{1,l}\circ\mb{b}_1\circ\mb{c}_1+\sum_{l=1}^{L_2}\mb{a}^{(1)}_{2,l}\circ\mb{a}^{(2)}_{2,l}\circ\mb{b}_2\circ\mb{c}_2+ \nonumber \\ 
    & & \cdots+\sum_{l=1}^{L_R}\mb{a}^{(1)}_{R,l}\circ\mb{a}^{(2)}_{R,l}\circ\mb{b}_R\circ\mb{c}_R,
    \label{eq:LL11expanded}
\end{eqnarray}
where the $\mb{c}_r$'s stand for the subject weights. Note that the BTD model results from expressing the $r$th spatial map vector in~\eqref{eq:CPD} as a sum of $L_r$ Kronecker products (equivalently, assuming that its matrix version is of rank $L_r$). 

One cannot guarantee the identifiability of the $\mb{A}^{(1)}_r,\mb{A}^{(2)}_r$ factors from their product (without additional constraints).\footnote{Nevertheless, this has been observed in our simulations, with both synthetic and real data~\cite{rkg21a,grk22}.} The model~\eqref{eq:LL1} is proved to be essentially unique, in the sense of having unique products $\mb{A}^{(1)}_r\mb{A}_r^{(2)\T}$ and vectors $\mb{b}_r$ with the usual scaling and permutation ambiguity, if $\mb{A}^{(1)},\mb{A}^{(2)}$ are of full column rank and there are no null or collinear columns $\mb{b}_r$~\cite{ldl08b}. This condition could also be expressed as the requirement of \emph{irreducibility}\cite{lj19b} for the block terms in~\eqref{eq:LL1expanded}. An analogous condition holds for the LL11 variant~\eqref{eq:LL11expanded}. \cite{t06} defines irreducibility as not being a mixture of independent sets of signals. 

Such a blind source separation approach is known as block component analysis (BCA)~\cite{ldl12} (also referred to as the block component model (BCM) in~\cite{vnhl12}). LL1 modeling has been thoroughly studied recently, giving rise to methods of \emph{joint model selection and computation}, that is, estimating the model order ($R$ and $L_r, r=1,2,\ldots,R$) and its parameters (including, for example, regularization-based~\cite{rkg21a} and Bayesian~\cite{grk22} methods). Given that CPD is a special case, these tools provide ways of determining the tensor rank as well. 

\section{Towards a ``TPIVA" Approach}
\label{eq:TIVA}

A possible extension to TPICA is brought out if one reconsiders the rank-1 terms of each sum in~\eqref{eq:LL1expanded}. Let the $\mb{b}$ vectors within each sum be not equal but ``similar" to each other. The similarity here might be meant also in statistical terms: the $\mb{b}$s within each sum term may be seen as dependent while those in different terms as independent. In the IVA parlance, each $\mb{b}_r$ is an $L_r$-variate source or source component vector (SCV). A new tensor decomposition model, in the sense of generalized CPD\footnote{By the way, this is another less well-known name for BTD~\cite{lb08}.}, would then result, where collinearity would be replaced by having ``soft" / ``dependence" links among sources~\cite{lahat21}. It is a critical question how such a BTD model would be defined, selected, and computed. The clustering of similar CPD terms, which has been used elsewhere (see, e.g., \cite{hzwlzwcn19} on clustering of rank-1 components for ICA dimensionality selection or~\cite{t06} and references therein), is relevant here (cf.~\cite{rkg21a} for a BTD literature review). Such a viewpoint of BTD brings IVA into the picture and generalizes the CPD model underlying TPICA. 

It is interesting to observe that each $K\times L_r$ matrix $\mb{\bar{B}}_r\triangleq\mb{b}_r\mb{1}_{L_r}^{\T}=\bmx{cccc} \mb{b}_r & \mb{b}_r & \cdots & \mb{b}_r \emx$ could more generally be replaced by a low rank (not simply rank-1) matrix,\footnote{Thanks to T. Adal{\i}, University of Maryland Baltimore County, for pointing out~\cite{sgbaylca23,gala23}.} generalizing BTD and serving in the role of a SCV. This model of ``shared" (in the sense of strongly dependent) components, formulated in the same way as $\mb{\bar{B}}_r$ but with noise added to randomize the collinearity relation, also appears in~\cite[Eq.~(2)]{sgbaylca23} for expressing shared sources within an SCV and would take the following form here:
\begin{equation}
    \mb{\bar{B}}_r=\mb{b}_r\mb{1}_{L_r}^{\T}+\mb{Z}_r
    \label{eq:Br1}
\end{equation}
The aim of~\cite{sgbaylca23} is to distinguish such SCVs from those with less strongly dependent sources (i.e., exhibiting higher variability within each SCV) since IVA is only well-suited to cope with the latter kind\footnote{It would overparameterize SCVs with strongly shared entries~\cite{sgbaylca23}.}. A more relaxed (with greater variability) model could represent $\mb{\bar{B}}_r$ as
\begin{equation}
    \mb{\bar{B}}_r=\mb{U}_r\mb{V}_r^{\T}+\mb{Z}_r,
    \label{eq:Br2}
\end{equation}
namely, allowing it to be more generally of low rank, not necessarily of rank~one~\cite[Eq.~(12)]{gala23}. An alternative (not necessarily independent) direction could be to consider the $\mb{b}_r$s for each $r$ as being ``coupled", in a way analogous to the way that soft coupling is defined and employed in coupled decomposition models (cf., e.g., \cite{ckplth22} and references therein).

\medskip

\noindent
\emph{Remark.} In an fMRI source unmixing scenario, with the $\mb{A}$s in BTD representing spatial factors, SCVs $\mb{\bar{B}}_r$s would comprise time courses. Thus, this would correspond to temporal BCA, in analogy with temporal ICA (tICA), whereas TPICA, as described in Section~\ref{sec:TPICA}, performs spatial ICA (sICA) instead~\cite{capp01,gc22}. 

However, neither the number of the blocks, $R$, nor the block sizes, $L_r$, is known \emph{a priori}. It is common in the BTD practice to take all $L_r$s equal, say to $L$. Moreover, it has been observed that results are rather insensitive to $L$ overestimation (see, e.g., \cite{ckmt19} and references therein). Even so, however, $R$ needs to be accurately estimated if one is to interpret the extracted components. It should be noted that these questions for the deterministic LL1 model have been recently given satisfactory answers (see, e.g., \cite{rkg21a,grk22}). An interplay between IVA and BTD, as suggested above, can also provide solutions for the completely blind IVA (or ISA) problem, namely with an \emph{a priori unknown number of multi-variate sources of unknown dimensionality} (cf.~\cite{sgbaylca23} and references therein). 
Of course, the above ideas can be extended to the LL11 model, which is more appropriate for, e.g., the multi-subject setting in fMRI analysis.

\section{Related Work}
\label{sec:related}

A little-known extension to TPICA is proposed in~\cite{gp08}, for the case of more than one group of subjects, having different time courses and subject scores. It is suggested to augment the mixing matrix $\mb{M}$ by stacking the mixing matrices for each of several $G$ groups one above another, as
\[
\mb{M}=\left[\begin{array}{c} \mb{C}_1 \odot \mb{B}_1 \\ \mb{C}_2\odot\mb{B}_2 \\ \vdots \\ \mb{C}_G\odot\mb{B}_G \end{array}\right]
\]
An extension more reminiscent of parallel ICA consists of applying TPICA per group as in the so-called linked ICA approach of~\cite{gbsw11}. 
\cite{uhtj11} on the other hand develops an extension of TPICA where the rather restrictive CPD model is replaced by its Tucker generalization. Another modification is found in the way the analysis is done. In contrast to TPICA, where ICA is performed first, it is the Tucker decomposition that is first computed here, followed by a rotation of the factor in the mode where independence is assumed so that independence (in fact uncorrelatedness) is enforced. The core tensor is then rotated accordingly to preserve the data fitting accuracy. In addition to the methodology proposed in~\cite{uhtj11}, it is also of interest to think of the analogies between the modes (space, time, meteorological variables) of the atmospheric science data analyzed therein and those of an fMRI analysis. 

TPICA is criticized in~\cite{vnhl12} for putting together the PICA and CPD models in an ad-hoc manner. \cite{vnhl12} proposes instead to combine these two tools in a more theoretically justified manner, joining their effects in a way that is reminiscent of the Joint Approximate Diagonalization of Eigematrices (JADE) method for ICA. The resulting so-called ICA-CPA method\footnote{CPA stands for Canonical Polyadic Analysis.} relies on the fact that the higher-order cumulant tensor of the sources must be diagonal and forces this by diagonalizing it with the aid of CPD-decomposed eigen-tensors. Interestingly, the TPIVA idea outlined above is also alluded to here, in the sense of block diagonalizing the cumulant tensor. A reference to the link between (symmetric) BTD and a source cumulant tensor of block diagonal structure was also made recently in~\cite{cl21}, which studied the problem of joint BTD model selection and computation through a block diagonal-core TKD point of view. The procedure developed relies on algebraic (SVD-type) (as is also the case in the method of~\cite{dd20}) and clustering steps. 

There is certainly a close connection between the above and the ideas of (joint) ISA and the concept of (i)reducibility of block terms thoroughly studied in~\cite{lj19a,lj19b,lahat21}. Nevertheless, it must be noted that the block decompositions appearing in those works are about the second-order statistics (covariances) of the data/sources. For the rest, however, there is certainly a lot of analogy with the cumulant block diagonalization mentioned previously. Joint ISA was given in~\cite{lj18} a double-coupled CPD (DC-CPD) interpretation, similar to the double-coupled CPD approach later proposed in~\cite{glcl18,glgl19}. A double-coupled matrix-tensor decomposition approach, applied in the data itself, not in its covariances, was also recently proposed in the context of fMRI and electroencephalography (EEG) fusion~\cite{ckplth22}. The close connection of joint block diagonalization of a set of matrices with BTD is pointed out in~\cite{nion11}.

\cite{lahacsa22} has recently compared IVA and a generalization of CPD, PARAFAC2,\footnote{PARAFAC2 relaxes the constraint of invariance (allowing, for example, different time courses per subject) imposed by CPD. In~\cite{ckmt19}, this was taken further, to an analogous extension of BTD, named BTD2.} in a multi-task fMRI fusion context. IVA was better when subject scores differed among datasets, while the tensorial method prevailed for scores that were proportional to each other. IVA was recently shown to be equivalent to DC-CPD under a piecewise stationary Gaussian model for the sources~\cite{rlcl22}. Again, DC-CPD was only better when the data did not follow the statistical model closely enough. 


The most recent (to the author's best knowledge) allusion to the idea proposed in this note appears in~\cite{luo23}. In Section~4.4, when comparing tensor decomposition and IVA, it is stressed that the former relies on shared deterministic information, which can be too restrictive in practice, while the latter is more flexible in this respect. This remark is in the same vein as that discussed here. \cite{luo23} only mentions CPD and TKD as candidate tensor models, however.

\section{Additional Work}
\label{sec:future}

Relying on a deterministic (tensor) model certainly relieves us from the risk of statistical model mismatch. Yet this does not benefit from the gains offered by exploiting the data statistics. Instead of taking one way or the other, it seems more natural and promising to try to exploit both at the same time. TPICA, despite its proven drawbacks, shows the way to such a hybrid approach and inspires the ``TPIVA" idea outlined here. 

The present note aims to briefly state the problem and outline the relevant ideas and possible directions for its solution. Beyond the basic ``TPIVA'' technique, online versions (e.g., \cite{kktcm21}) will also be developed and applied in, e.g., dynamic neuroimaging~\cite{ca16}. To that end, recent work on online BTD that both estimates and tracks the model ranks\cite{rkg23} will prove very useful. Scalable schemes could be developed for large-scale related scenarios, as in, e.g., \cite{gsaca23,sgbaylca23}.

It should be stressed that the possible applications go far beyond functional neuroimaging and include analogous problems in remote sensing (hyperspectral imaging~\cite{grk22}), radar~\cite{lima2020}, and communications~\cite{lgl22}, among many others.

\bibliographystyle{IEEEtran}
\bibliography{IEEEabrv,references}

@STRING{IEEE_J_STSP       = "{IEEE} J. Sel. Topics Signal Process."}

@STRING{IEEE_J_SPL        = "{IEEE} Signal Process. Lett."}

@STRING{IEEE_J_SP         = "{IEEE} Trans. Signal Process."}

@STRING{IEEE_J_MI         = "{IEEE} Trans. Med. Imag."}

@STRING{IEEE_M_SP         = "{IEEE} Signal Process. Mag."}

@article{l08,
author = "M. A. Lindquist",
title = "The statistical analysis of {fMRI} data",
journal = "Stat. Sci.",
volume = 23,
number = 4,
pages = "439--464",
year = 2008
}

@article{w12,
author = "M. W. Woolrich",
title = {{Bayesian inference in fMRI}},
journal = "NeuroImage",
volume = 62,
pages = "801--810",
year = 2012
}

@article{zlz13,
author = "H. Zhou and L. Li and H. Zhu",
title = "Tensor Regression with Applications in Neuroimaging Data Analysis",
journal = "J. Am. Stat. Assoc.",
volume = 108,
number = 502,
pages = "540–-552",
year = 2013
}

@article{ar04,
author = "Anders H. Andersen and William S. Rayens",
title = "{Structure-seeking multilinear methods for the analysis of fMRI data}",
journal = "NeuroImage",
volume = 22,
number = 2,
pages = "728--739",
year = 2004
}

@article{bs04,
author = "C. F. Beckmann and S. M. Smith",
title = "Probabilistic Independent Component Analysis
for Functional Magnetic Resonance Imaging",
journal = IEEE_J_MI,
volume = 23,
number = 2,
pages = "137--152",
month = feb,
year = 2004
}

@article{bs05,
author = "C. F. Beckmann and S. M. Smith",
title = "Tensorial extensions of independent component analysis for multisubject {fMRI} analysis",
journal = "NeuroImage",
volume = 25,
pages = "294--311",
year = 2005
}

@article{uhtj11,
author = "Steffen Unkel and others",
title = "Independent Component Analysis for Three-Way Data With an Application From Atmospheric Science",
journal = "J. Agric. Biol. Env. Stat.",
volume = 16,
number = 3,
pages = "319–-338",
year = 2011
}

@article{gbsw11,
author = "Adrian R. Groves and others",
title = "Linked independent component analysis for multimodal data fusion",
journal = "NeuroImage",
volume = 54,
pages = "2198–2217",
year = 2011
}

@article{vnhl12,
author = "M. {De~Vos} and others",
title = "A combination of parallel factor and independent component analysis",
journal = "Signal Process.",
volume = 92,
number = 12,
pages = "2990--2999",
month = dec,
year = 2012
}

@article{gp08,
author = "Ying Guo and Giuseppe Pagnoni",
title = "{A unified framework for group independent component analysis for multi-subject fMRI data}",
journal = "NeuroImage",
volume = 42,
number = 3,
pages = "1078--1093",
year = 2008
}

@techreport{s07,
    author    = "A. Stegeman",
    title     = {{Comparing Independent Component Analysis and the Parafac model for artificial multi-subject fMRI data}},
    institution = "University of Groningen",
    year      = "2007",
    month    = feb
}

@inproceedings{fgc18,
    author = "Rodrigo Cabral Farias and Jos\'{e} Henrique {de~Morais~Goulart} and Pierre Comon",
    title = "Coherence Constrained Alternating Least Squares",
    booktitle = "Proc.~{EUSIPCO-2018}",
    address = "Rome, Italy",
    month = sep,
    year = 2018
}

@article{hh13,
author = "N. E. Helwig and S. Hong",
title = {{A critique of Tensor Probabilistic Independent Component Analysis: Implications and recommendations for multi-subject fMRI data analysis}},
journal = "J. Neurosci. Meth.",
volume = 213,
number = 2,
pages = "263--273",
month = mar,
year = 2013
}

@article{hzwlzwcn19,
author = "G. Hu and others",
title = "Tensor clustering on outer-product of coefficient and component matrices of independent component analysis for reliable functional magnetic resonance imaging data decomposition",
journal = "J. Neurosci. Meth.",
volume = 325,
number = 108359,
month = sep,
year = 2019
}

@article{capp01,
author = "V. D. Calhoun and others",
title = {{Spatial and temporal independent component analysis of functional MRI data containing a pair of task‐related waveforms}},
journal = "Hum. Brain Mapp.",
volume = 13,
number = 1,
pages = "43--53",
month = may,
year = 2001
}

@article{gc22,
author = "Ali M. Golestani and J. Jean Chen",
title = {{Performance of temporal and spatial independent component analysis in identifying and removing low-frequency physiological and motion effects in resting-state fMRI}},
journal = "Front. Neurosci.",
volume = 16,
month = jun,
year = 2022
}

@article{ca16,
author = "V. D. Calhoun and T. Adal{\i}",
title = "{Time-varying brain connectivity in fMRI data: Whole-brain data-driven approaches for capturing and characterizing dynamic states}",
journal = IEEE_M_SP,
volume = 33,
number = 3,
pages = "52--66",
month = may,
year = 2016
}

@inproceedings{gsaca23,
    author = "Ben Gabrielson and others",
    title = {{Independent vector analysis with multivariate Gaussian model: A scalable method by multilinear regression}},
    booktitle = "Proc.~{ICASSP-2023}",
    address = "Rhodes, Greece",
    month = jun,
    year = 2023
}

@article{gala23,
author = "B. Gabrielson and others",
title = "An efficient analytic solution for joint blind source separation",
journal = IEEE_J_SP,
volume = 72,
pages = "2436--2449",
month = may,
year = 2024
}

@article{sgbaylca23,
author = "M. Sun and others",
title = {{A scalable approach to independent vector analysis by shared subspace separation for multi-subject fMRI analysis}},
journal = "Sensors",
volume = 23,
doi = "https://doi.org/10.3390/s23115333",
year = 2023
}

@article{akasca22,
author = "T. Adal{\i} and others",
title = "Reproducibility in Matrix and Tensor Decompositions: Focus on model match, interpretability, and uniqueness",
journal = IEEE_M_SP,
volume = 39,
number = 4,
pages = "8--24",
month = jul,
year = 2022
}

@article{mcm17,
author = "Kristoffer H. Madsen and Nathan W. Churchill and Morten M{\o}rup",
title = "{Quantifying functional connectivity in multi‐subject fMRI data using component models}",
journal = "Hum. Brain Mapp.",
volume = 38,
number = 2,
pages = "882–-899",
year = 2017
}

@inproceedings{t06,
author = "F. J. Theis",
title = "Towards a general independent subspace analysis",
booktitle = "Proc.~{NIPS-2006}",
address = "Vancouver, BC, Canada",
month = dec,
year = 2006
}

@inproceedings{lj19b,
author = "D. Lahat and C. Jutten",
title = "Tensor and Coupled Decompositions in Block Terms: Uniqueness and Irreducibility",
booktitle = "Proc.~{SPARS-2019}",
address = "Toulouse, France",
month = jul,
year = 2019
}

@inproceedings{lj18,
author = "D. Lahat and C. Jutten",
title = "A New Link Between Joint Blind Source Separation Using Second Order Statistics and the Canonical Polyadic Decomposition",
booktitle = "Proc.~{LVA/ICA-2018}",
address = "Guilford, UK",
month = jul,
year = 2018
}

@article{nion11,
author = "D. Nion",
title = "A Tensor Framework for Nonunitary Joint Block Diagonalization",
journal = IEEE_J_SP,
volume = 59, 
number = 10, 
pages = "4585--4594",
month = oct,
year = 2011
}

@inproceedings{kll06,
author = "T. Kim and I. Lee and T.-W. Lee",
title = "Independent vector analysis: Definition and algorithms",
booktitle = "Proc.~{ACSSC-2006}",
address = "Pacific Grove, CA",
month = "29~Oct.--1~Nov.",
year = 2006
}

@article{lljy08,
author = "J.-H. Lee and T.-W. Lee and F. A. Jolesz and S.-S. Yoo",
title = "{Independent vector analysis (IVA): Multivariate approach for fMRI group study}",
journal = "NeuroImage",
volume = 40,
pages = "86--109",
year = 2008
}

@article{luo23,
author = "Z. Luo",
title = "Independent vector analysis: Model, applications, challenges",
journal = "Pattern Recognit.",
volume = 138,
number = 109376,
year = 2023
}

@inproceedings{lahacsa22,
author = "Isabell Lehmann and others",
title = {{Multi-task fMRI data fusion using IVA and PARAFAC2}},
booktitle = "Proc.~{ICASSP-2022}",
address = "Singapore",
month = may,
year = 2022
}

@article{lj19a,
author = "D. Lahat and C. Jutten",
title = "Joint independent subspace analysis: Uniqueness and identifiability",
journal = IEEE_J_SP,
volume = 67,
number = 3,
pages = "684--699",
month = feb,
year = 2019
}

@misc{lahat21,
author = "D. Lahat",
title = "Beyond independent component analysis: Subspace, coupled, and block decompositions",
howpublished = {{The Brain Space Initiative Talk Series}},
url = "https://www.youtube.com/watch?v=WOh2zvj4aWU\&t=2s",
year = 2021
}

@article{kktcm21,
author = "Z. Koldovsk{\'y} and others",
title = "Dynamic Independent Component/Vector Analysis: Time-Variant Linear Mixtures Separable by Time-Invariant Beamformers",
journal = IEEE_J_SP,
volume = 69,
pages = " 2158 - 2173",
month = mar,
year = 2021
}

@article{lima2020,
author = "D. V. {de~Lima} and others",
title = {{Robust tensor-based techniques for antenna array-based GNSS receivers in scenarios with highly correlated multipath components}},
journal = "Digit. Signal Process.",
volume = 101,
month = jun,
year = 2020
}

@article{lgl22,
author = "Z. Luo and R. Guo and C. Li",
title = "Independent vector analysis for blindly deconvolving digital modulated communication signals",
journal = "Electronics",
volume = 11,
number = 1460,
year = 2022
}

@article{rkg21a, 
author={A. A. Rontogiannis and E. Kofidis and P. V. Giampouras},  
title="Block-Term Tensor Decomposition: Model Selection and Computation",
journal= IEEE_J_STSP, 
volume = 15,
number = 3, 
pages = "464--475",
month = apr,
year = 2021
}

@article{grk22,
author = "P. V. Giampouras and A. A. Rontogiannis and E. Kofidis",
title = "{Block-term tensor decomposition model selection and computation: The Bayesian way}",
journal = IEEE_J_SP,
volume = 70,
pages = "1704--1717",
year = "2022"
}

@article{rkg23,
author = "A. A. Rontogiannis and E. Kofidis and P. V. Giampouras",
title = "Online Rank-Revealing Block-Term Tensor Decomposition",
journal = "Signal Process.",
volume = 212, 
number = 109126,
month = nov,
year = 2023
}

@article{s12,
author = "A. Stegeman",
title = "{CANDECOMP/PARAFAC}: From diverging components to a decomposition in block terms",
journal = {{SIAM J. Matrix Anal. Appl.}},
volume = 33,
number = 2,
pages = "291–-316",
year = 2012
}

@article{bdl17,
author = "M. Bouss{\'e} and O. Debals and L. {De~Lathauwer}",
title = "A Tensor-Based Method for Large-Scale Blind Source Separation Using Segmentation",
journal = IEEE_J_SP,
volume = 65,
number = 2,
pages = "346--358",
month = jan,
year = 2017
}

@article{lb08,
author = "L. {De~Lathauwer} and A. {De~Baynast}",
title = {{Blind deconvolution of DS-CDMA signals by means of decomposition in rank-$(1, L, L)$ terms}},
journal = IEEE_J_SP,
volume = 56,
number = 4,
pages = "1562--1571",
month = apr,
year = 2008
}

@article{ldl08a,
author = "L. {De~Lathauwer}",
title = "Decompositions of a higher-order tensor in block terms --- {P}art~{I}: Lemmas for partitioned matrices",
journal = "{SIAM} J. Matrix Anal. Appl.",
volume = 30,
number = 3,
pages = "1022--1032",
year = 2008
}

@article{ldl08b,
author = "L. {De~Lathauwer}",
title = "Decompositions of a higher-order tensor in block terms --- {P}art~{II}: Definitions and uniqueness",
journal = "{SIAM} J. Matrix Anal. Appl.",
volume = 30,
number = 3,
pages = "1033--1066",
year = 2008
}

@article{ldl08c,
author = "L. {De~Lathauwer} and D. Nion",
title = "Decompositions of a higher-order tensor in block terms --- {P}art~{III}: Alternating least squares algorithms",
journal = "{SIAM} J. Matrix Anal. Appl.",
volume = 30,
number = 3,
pages = "1067--1083",
year = 2008
}

@inproceedings{ldl12,
author = "L. {De~Lathauwer}",
title = "Block component analysis: a new concept for blind source separation",
booktitle = "Proc.~{LVA/ICA-2012}",
month = mar,
year = 2012,
address = "Tel Aviv, Israel"
}

@article{ckmt19,
author = "C. Chatzichristos and others",
title = "Blind {fMRI} source unmixing via higher-order tensor decompositions",
journal = "J. Neurosci. Meth.",
volume = 315,
pages = "17--47",
month = mar,
year = 2019
}

@inproceedings{cvktlv19,
author = "C. Chatzichristos and others",
title = "Tensor-based Blind {fMRI} Source Separation Without the {Gaussian} Noise Assumption — {A} $\beta$-Divergence Approach",
booktitle = "Proc.~{GlobalSIP-2019}",
address = "Ottawa, Canada",
month = nov,
year = 2019
}

@article{ckplth22,
author = "C. Chatzichristos and others", 
title = {{Early soft and flexible fusion of EEG and fMRI via Double CMTF for multi-subject group analysis}},
journal = "Hum. Brain Mapp.",
volume = 43,
number = 4,
pages = "1231–-1255", 
year = 2022,
doi = "https://doi.org/10.1002/hbm.25717"
}

@article{sdfhpf17,
author = "N. D. Sidiropoulos and others",
title = "Tensor decomposition for signal processing and machine learning",
journal = IEEE_J_SP,
volume = 65,
number = 13,
pages = "3551--3582",
month = jul,
year = 2017
}

@article{dd20,
author = "I. Domanov and L. {De~Lathauwer}",
title = "On Uniqueness and Computation of the Decomposition of a Tensor into Multilinear Rank-{$(1,L_r,L_r)$} Terms",
journal = "SIAM J. Matrix Anal. Appl.",
volume = 41,
number = 2,
year = 2020
}

@inproceedings{cl21,
author = "Yunfeng Cai and Ping Li",
title = "A Blind Block Term Decomposition of High Order Tensors",
booktitle = "Proc.~{AAAI-2021}",
month = feb,
year = 2021
}

@article{glcl18,
author = "Xiao-Feng Gong and others",
title = "Double Coupled Canonical Polyadic Decomposition for Joint Blind Source Separation",
journal = IEEE_J_SP,
volume = 66,
number = 13,
pages = "3475--3490",
month = jul,
year = 2018
}

@article{glgl19,
author = "Xiao-Feng Gong and others",
title = "Double coupled canonical polyadic decomposition of third-order tensors:
Algebraic algorithm and relaxed uniqueness conditions",
journal = "Signal Process.: Image Commun.",
volume = 73,
pages = "22--36",
year = 2019
}

@article{rlcl22,
author = "Haoxin Ruan and others",
title = "An Explicit Connection Between Independent Vector Analysis and Tensor Decomposition in Blind
Source Separation",
journal = IEEE_J_SPL,
volume = 29,
pages = "1277--1281",
month = may,
year = 2022
}

\end{document}